\documentclass{elsart}
\usepackage{natbib}
\usepackage{graphicx}
\usepackage{amssymb}
\usepackage{amsmath}
\usepackage{bm}

\def\url#1{{\ttfamily\def\/{/\discretionary{}{}{}}#1}}
\def\bibcode#1{}

\newcommand\ba{\begin{eqnarray}}
\newcommand\ea{\end{eqnarray}}
\newcommand\eq{\begin{equation}}
\newcommand\en{\end{equation}}

\newcommand\bbr{\vec{ r}}
\newcommand\bbk{\vec{ k}}
\newcommand\bbK{\vec{K}}
\newcommand\vth{\vec{\theta}}
\newcommand\sgn{\rm sgn}

\begin{document}
\begin{frontmatter}
\title{Power spectrum and correlation function errors:
Poisson vs. Gaussian shot noise}
\author{J. D. Cohn\thanksref{jdcemail}}
\address{Space Sciences Lab and  Theoretical Astrophysics Center \\
University of California, Berkeley, CA 94720-3411, USA }
\thanks[jdcemail]{jcohn@astro.berkeley.edu}
\begin{abstract}
Poisson distributed shot noise is normally considered in
the Gaussian limit in cosmology.  However, if the shot noise is large enough 
and the correlation function/power spectrum conspires, the Gaussian 
approximation mis-estimates the errors and their covariance 
significantly.   The power spectrum, even for initially Gaussian densities,
acquires cross correlations which can be large, while the change in 
the correlation function error matrix is diagonal except at zero separation. 
 Two 
and three dimensional power law correlation function and power spectrum
examples are given.  These corrections appear to have a large effect
when applied to galaxy clusters, e.g. for SZ selected galaxy clusters in 2 
dimensions.  This can increase the error estimates for cosmological parameter 
estimation and consequently affect survey strategies, as the
corrections are minimized for surveys 
which are deep and narrow rather than wide and shallow.
In addition, a rewriting of the error matrix for the power 
spectrum/correlation 
function is given which eliminates most of the Bessel function dependence (in
two dimensions) and all of it (in three dimensions), which makes the
calculation of the error matrix more tractable.  This applies even when the
shot noise is in the (usual) Gaussian limit.  
\end{abstract}
\end{frontmatter}

In cosmology, continuous density fields are used to describe
distributions of discrete objects such as galaxy clusters.
The actual discrete number counts are taken to be a sampling 
of this continuous distribution resulting in the appearance of shot noise.
Usually the shot noise is taken to have a Gaussian distribution,
considered as the large number approximation to an underlying
Poisson distribution (Layzer \cite{Lay56}, see the textbook by Peebles
\cite{Pee80} for a review). 
The interest here is when the Gaussian approximation for
the shot noise breaks down.   This paper considers in detail the specific 
case where the shot noise distribution is in fact Poisson.\footnote{It
has been shown that Poisson sampling is accurate for dark matter halos 
in numerical simulations in regions where the density is not too high 
(Casas-Miranda et al \cite{Casetal02}); the transition density
depends upon the dark halo mass of interest.}
It is seen that the Poisson nature of the shot noise can produce significant
modifications of the Gaussian shot noise error matrix of the
correlation function and power spectrum.

The full Poisson power spectrum error matrix 
was calculated by Meiksin \& White \cite{MeiWhi99}.  In this paper
it is angle averaged, compared to usual expressions, and
applied to 2 and 3 dimensional power law examples comparable to
SZ selected galaxy cluster surveys.  The effect on the explicit
two dimensional SZ selected galaxy cluster power spectrum is also found
and is seen to be large for some realistic survey parameters.
These increases to the error matrix have consequences both for cosmological 
parameter estimates
and for designing surveys (the additions only appear if
the shot noise is big enough, which can be alleviated by going
deeper for an SZ selected survey).  

Section one has definitions
and the two dimensional angle averaging of the Meiksin \& White \cite{MeiWhi99} 
result.  Three dimensions are considered
in section two, and section three discusses a rewriting of the error matrix
which minimizes the appearance of Bessel functions and their instabilities.
This rewriting is also useful for the Gaussian shot noise case.
In section four, two and three dimensional power law examples
are given, as well as an application to a two dimensional SZ selected galaxy 
cluster power spectrum.  Section five concludes.

The rarity of galaxy clusters and the possible need for Poisson shot
noise has been recognized before, e.g. in Lima \& Hu 
\cite{LimHu04} the full Poisson probability function is used to
obtain constraints from cluster number counts in conjunction with
their scatter.

\section{The effect: two dimensions}
One can characterize a two dimensional density distribution via
its correlation function
\eq
\langle \hat{\delta}(\bbr) \hat{\delta}(\bbr+\vth) \rangle
\end{equation}
where $\vth$ refers to the two dimensional angular separation $(\theta,\phi)$
and the hat denotes an operator.
There are actually two averages in the above expression for discrete
objects, as
the density distribution is treated as a continuous
field sampled from the underlying discrete distribution, here the sampling
is taken to be Poisson.  The first
average is over this Poisson sampling, and the second is a sample
average, usually assumed to be the same as volume average.  
The Poisson and sampling average of two density perturbations 
gives\footnote{For more detailed discussion in Fourier space for the
conventions used here see, e.g., \cite{MeiWhi99}.}
\eq
\langle \hat{\delta}(\bbr) \hat{\delta}(\bbr+\vth) \rangle
= w(\vth) + \frac{\delta_D(\vth)}{\mathcal N}
\end{equation}
where
$\delta_D(\vth)$ is a two-dimensional Dirac delta function, and
${\mathcal N}$ is the number density of objects.   The
angular correlation function
$w(\vth)$ is defined as the probability above random for finding
two objects with one in area $\Omega_1$ and the other in area $\Omega_2$
\eq
 P_{12} = {\mathcal N}^2(1 + w(\vth)) d \Omega_1 d \Omega_2
\end{equation}

The operator
$\hat{w}$ is usually defined with the shot noise subtracted out:
\eq
\hat{w}(\vth) = \hat{\delta}(\bbr) \hat{\delta}(\bbr+\vth) 
-\frac{\delta_D(\vth)}{\mathcal N}
\end{equation}
so that
$\langle \hat{w}(\vth) \rangle = w(\vth)$.

In practice the measured correlation function
will also depend on window functions of the survey $W(\vth)$,
\eq
\begin{array}{l}
\hat{w}(\vth) =\frac{1}{A(\vth)} \int d^2 \theta' 
\int d^2 \theta'' W(\vth') W(\vth'') \hat{\delta}(\vth') \hat{\delta}(\vth'')
\delta_D (\vth'-\vth''-\vth)
-\frac{\delta_D(\vth)}{\mathcal N} \\
A(\vth) = \int d^2 \theta' d^2 \theta''  W(\vth') W(\vth'') 
\delta_D (\vth'-\vth''-\vth) \; .
\end{array}
\end{equation}
In the work
here the window function is taken to be one inside the region of the 
survey and zero elsewhere.  This simple case illustrates the effect of
interest for this paper, however in practice more complicated window functions
do arise and can introduce additional subtleties.  For more detailed 
discussion of window functions and errors see Eisenstein \& Zaldarriaga 
\cite{EisZal01}.

There is also the Fourier transform of $w(\vth)$, the power spectrum, 
\eq
P_2(\bbK)
= \int d^2 \theta e^{-i \bbK \cdot \vth}w(\vth) \; ,
\end{equation}
here $\bbK = (K,\phi_k)$.  The operator $\hat{P}_2(\bbK)$
also has the shot noise subtracted out (in this case a constant).
The operators $\hat{\delta}(\vth),\hat{w}(\vth)$ are dimensionless, while 
$\hat{P}_2(\bbK)$ is in terms of steradians.

As isotropy is usually assumed in $\bbK$ and $\vth$ space (this can
mean neglecting boundary effects in particular), both $w$ and $P_2$
are usually angle averaged:
\eq
\begin{array}{l}
w(\theta) = \frac{1}{2 \pi} \int d \phi w(\vth) \\
P_2(K) = \frac{1}{2 \pi} \int d \phi_K P_2(\bbK)
\end{array}
\end{equation}

The error matrix of the correlation function and power spectrum
can be calculated by going back to the definitions of $\hat{w}(\vth)$
and $\hat{P}(\bbK)$ in terms of the density.
Meiksin \& White \cite{MeiWhi99} calculated the Poisson contribution
to the error matrix\footnote{The two dimensional version is shown below,
i.e. take $V \rightarrow \Omega$ in their result.} 
in the power spectrum using counts in cells (Peebles,
\cite{Pee80} \S 36).  For Gaussian densities $\hat{\delta}$
the power spectrum error matrix is 
\eq
\label{powervar}
\begin{array}{ll}
\langle \hat{P}_2(\bbK) \hat{P}_2(\bbK') \rangle -
\langle \hat{P}_2(\bbK) \rangle \langle \hat{P}_2(\bbK') \rangle 
 &=
(P_2(\bbK) + \frac{1}{\mathcal N})^2 (\delta_{\bbK,\bbK'} + \delta_{\bbK,-\bbK'}) \\
& \; + \frac{1}{{\mathcal N}^2 \Omega}( P_2(|\bbK + \bbK'|)
+ P_2(|\bbK-\bbK'|)+2P_2(\bbK) +2 P_2(\bbK')) \\ 
& \; + \frac{1}{{\mathcal N}^3 \Omega} \\
\end{array}
\end{equation}
Here (and later in the text) $\delta_{ab}$ refers to a Kronecker $\delta$
function.
If the densities themselves are
non-Gaussian, there are additional terms depending upon
the three and four point functions,
\eq
\begin{array}{l}
\frac{1}{\Omega}T(\bbK,-\bbK,\bbK',-\bbK')
 +
\frac{1}{{\mathcal N} \Omega} [B(\bbK,-\bbK,0) + B(0,\bbK',-\bbK')\\
 +
B(\bbK+\bbK',-\bbK,-\bbK') + B(\bbK-\bbK',-\bbK,\bbK')
 +
B(\bbK,\bbK'-\bbK,-\bbK') +B(\bbK,-\bbK-\bbK',\bbK')] 
\end{array}
\end{equation}
which are Fourier transforms of the spatial four and three point functions
$\eta$ and $\zeta$ respectively.
These additional terms can be angle averaged (and Fourier transformed)
straightforwardly and thus are not shown in the following.
They are needed however if the density distribution is non-Gaussian.

There is also an additional term $\frac{1}{{\mathcal N}^3 \Omega} P_2(0)$,
which may be nonvanishing for some choices of window function.
As it is a constant it just modifies
\eq
\frac{1}{{\mathcal N}^3 \Omega} \rightarrow 
\frac{1}{{\mathcal N}^3 \Omega} (1 + {\mathcal N} P_2(0))
\end{equation} 
in the above and in its averages/limits below.  The same multiplicative
factor ($1 +{\mathcal N} P_2(0)$) (if different from one)
occurs in the the last term in the error matrix for the two
dimensional correlation function.  For three dimensions the last
term is instead modified by $(1 + {\mathcal N}P(0))$.  
It will not be displayed in the
following but can be added in directly if non-zero.\footnote{I thank the
anonymous referee for making this point.}

To convert this expression to errors and covariances for
the angle averaged correlation function
and power spectrum, first take the continuum limit.  As
\eq
\sum_k \rightarrow \frac{\Omega}{(2 \pi)^2} \int d^2 K \; 
\; \; \mathrm {and } \; \; \; 
\sum_k \delta_{\bbK,\bbK'} = 1,
\end{equation}
one has the correspondence
\eq
\delta_{\bbK,\bbK'} \rightarrow 
\frac{(2 \pi)^2}{\Omega} \delta_D({\bbK- \bbK'}) \;
\end{equation}
and similarly for $\delta_{\bbK,-\bbK'}$.
Thus the two dimensional power spectrum error matrix is
\eq
\label{2dperror}
\begin{array}{ll}
\langle \hat{P}_2(\bbK) \hat{P}_2(\bbK') \rangle -
\langle \hat{P}_2(\bbK) \rangle \langle \hat{P}_2(\bbK') \rangle 
&=
(P_2(\bbK) + \frac{1}{\mathcal N})^2\frac{(2 \pi)^2}{\Omega} [
\delta_D({\bbK- \bbK'}) + \delta_D({\bbK+ \bbK'}) ]
\\ &+
 \frac{1}{{\mathcal N}^2 \Omega}[2P_2(\bbK)+2P_2(\bbK')+P_2(|\bbK + \bbK'|) +
P_2(|\bbK - \bbK'|)] \\
& + \frac{1}{{\mathcal N}^3 \Omega} 
\end{array}
\end{equation}
The Fourier transform of this is the variance of the correlation
function $w(\vth)$:
\eq
\label{2dwerror}
\begin{array}{ll}
\langle \hat{w}(\vth) \hat{w}(\vth') \rangle -
\langle \hat{w}(\vth)\rangle \langle \hat{w}(\vth') \rangle
&=\frac{1}{\Omega}\int d^2 \theta''(w(\vth-\vth'-\vth'') + w(\vth + \vth' -\vth''))w(\vth'') \\
&+
\frac{1}{\Omega{\mathcal N}} [2 w(\vth+\vth') + 2 w(\vth -\vth')]
\\ &+
\frac{1}{\Omega{\mathcal N}^2}[(\delta_D(\vth-\vth')+
\delta_D(\vth+\vth'))(1 + w(\vth))
\\ &\; \; \; \; \; \; +  
2 w(\vth) \delta_D(\vth') 
+ 2 w(\vth') \delta_D(\vth)]
\\ & + \frac{1}{\Omega{\mathcal N}^3} \delta_D(\vth) 
\delta_D(\vth')
\end{array}
\end{equation}

These correlation functions and power spectra are not angle averaged,
i.e., they depend upon $\vth$ and $\bbK$ rather than $\theta,K$.  To get
the more familiar functions of only $\theta, \theta'$ and $K,K'$ requires an 
average over $\phi,\phi'$ and $\phi_k,\phi_k'$ respectively.  
(In practice
this means ignoring boundary effects, see Eisenstein and Zaldariagga
\cite{EisZal01} for discussion.)  The correlation function and
power spectrum are considered in turn below.  The error matrix
for each is angle averaged and either
binned or discretized.   Then each is compared with previous Gaussian 
expressions in the literature to highlight when and how the additional
Poisson-only (or ``non-Gaussian'') terms are significant.

For the angular correlation function, angle averaging requires the quantities
\eq
\frac{1}{2 \pi}\int d\phi \delta_D(\vth-\vth') =\frac{1}{2 \pi} \int d\phi 
\frac{ \delta_D(\theta -\theta')}
{\theta}
\delta_D (\phi - \phi') = \frac{1}{2 \pi}\frac{\delta_D (\theta-\theta')}{\theta} 
\end{equation}
and 
\eq
\frac{1}{2 \pi}\int d\phi \delta_D(\vth+\vth') =
\frac{1}{2 \pi} \int d\phi \frac{ \delta_D(\theta -\theta')}
{\theta}
\delta_D (\phi - \phi'\pm \pi  ) = 
\frac{1}{2 \pi}\frac{\delta_D (\theta-\theta')}{\theta}
\end{equation}
and likewise
\eq
\frac{1}{2 \pi}\int d\phi \delta_D(\vth) = \frac{1}{2 \pi \theta} \delta_D(\theta)
\end{equation}
These relations allow one to get
 the angle average of all the terms except those
containing two $w$'s.  For those, go to Fourier space to get
\eq
\label{fourspace}
\begin{array}{ll}
 \frac{1}{(2\pi)^2} \int d\phi d \phi' 
\int d^2 \vth''w(\vth-\vth'-\vth'') w(\vth'')
 &=
\ \frac{1}{(2\pi)^2} \int d\phi d \phi' 
\int \frac{d^2 K}{(2 \pi)^2}
e^{i \bbK \cdot (\vth-\vth')} P_2^2(K) \\
\; \; \; \; 
&=\int \frac{K d K}{2 \pi} J_0(K \theta) J_0(K \theta') P_2^2(K)
\\ \\
 \frac{1}{(2\pi)^2} \int d\phi d \phi'
\int d^2 \vth'' 
w(\vth + \vth' -\vth'')w(\vth'') &=
\int \frac{K d K}{2 \pi} J_0(K \theta) J_0(K \theta') P_2^2(K)\\ 
\\
\frac{1}{(2\pi)^2} \int d\phi d \phi'w(\vth+\vth')&=
\frac{1}{(2\pi)^2} \int d\phi d \phi'w(\vth-\vth') \\
&=
\int \frac{K d K}{2 \pi} J_0(K \theta) J_0(K \theta') P_2(K) \\
\end{array}
\end{equation}

Putting it all together gives
\eq
\label{correr1}
\begin{array}{ll}
\langle \hat{w}(\theta)\hat{w}(\theta') \rangle
-&\langle \hat{w}(\theta) \rangle \langle \hat{w}(\theta')) \rangle
\\ 
& =\frac{2}{\Omega}\int \frac{K d K}{2 \pi} J_0(K \theta) J_0(K \theta') 
[P_2^2(K) + \frac{2}{{\mathcal N}} P_2(K)] 
 +\frac{2}{\Omega{\mathcal N}^2} 
\frac{\delta_D(\theta-\theta')}{2 \pi \theta} \\ 
& \; +\frac{1}{\Omega{\mathcal N}^2} 
[2 w(\theta)
\frac{\delta_D(\theta-\theta')}{2 \pi \theta} +
 2 w(\theta) \frac{\delta_D(\theta')}{2 \pi \theta'} 
+2 w(\theta') \frac{\delta_D(\theta)}{2 \pi \theta}] \\
& \; +
\frac{1}{\Omega{\mathcal N}^3}
\frac{1}{(2 \pi \theta)(2 \pi \theta') } \delta_D(\theta)
\delta_D(\theta')
\end{array}
\end{equation}

Often the error in $w$ is given in terms of $n_p$, the number of
pairs.  To connect with this expression, binning is needed,
i.e. one observes not at just one angle but
in a shell of width $\delta \Omega$.
This means $w(\theta)$ is not measured at a point but
smeared out\footnote{I thank R. Sheth for correcting an error in this,
in the following expression, and equation 26.} 
over a region of size $\Delta \theta$:
\begin{equation}
\hat{w}(\theta) \rightarrow \frac{
\int_{-\Delta \theta/2}^{\Delta \theta/2} 
\hat{w}(\theta + \Delta \theta') (\theta + \Delta \theta') 
d (\Delta \theta')}
{\int_{-\Delta \theta/2}^{\Delta \theta/2} (\theta + \Delta \theta') 
d (\Delta \theta')}  \equiv 
\hat{w}_{\Delta \theta}(\theta)\; .
\end{equation}
In practice this means that
in equation \ref{correr1} one must replace
\eq
\begin{array}{rl}
J_0(K \theta) &\rightarrow \frac{1}{\theta \Delta \theta}
\int_{-\Delta \theta/2}^{\Delta \theta/2} 
J_0(K (\theta + \Delta \theta')) (\theta + \Delta \theta') 
d (\Delta \theta') \\
&= \frac{1}{K \theta \Delta \theta } ((\theta_+)J_1(K \theta_+) -(\theta_-)J_1(K \theta_-)) \\
&\equiv 
J_{0,\Delta \theta}(K \theta) 
\\
\theta_\pm & = \theta \pm \Delta \theta/2 \\
\end{array}
\end{equation}
as well as the explicit functions of $w(\theta), w(\theta')$ on
the right hand side.  The continuous $\delta_D(\theta-\theta')$
functions are effectively replaced by $\delta_{\theta,\theta'}/\Delta \theta$. 
Recognizing
$2 \pi \theta \Delta \theta =\delta \Omega $ gives
\eq
\begin{array}{ll}
\langle \hat{w}_{\Delta \theta}(\theta)\hat{w}_{\Delta \theta}(\theta') \rangle \; \;
-&\langle \hat{w}_{\Delta \theta}(\theta) \rangle \langle \hat{w}_{\Delta \theta}(\theta')) \rangle
 \\ 
&
=\frac{2}{\Omega}\int \frac{K d K}{2 \pi} J_{0,\Delta \theta}(K \theta) 
J_{0,\Delta \theta}(K \theta') [P_2^2(K) + \frac{2}{\mathcal N} P_2(K)]
+\frac{2}{\Omega{\mathcal N}^2}
\frac{\delta_{\theta,\theta'}}{\delta \Omega} 
\\
&
\; +\frac{1}{\Omega{\mathcal N}^2}
[2 w_{\Delta \theta}(\theta)
\frac{\delta_{\theta,\theta'}}{\delta \Omega} +
2 w_{\Delta \theta} (\theta) \frac{\delta_{\theta',0}}{\delta \Omega'} 
+2 w_{\Delta \theta} (\theta') \frac{\delta_{\theta,0}}{\delta \Omega}] \\
& \; +
\frac{1}{\Omega{\mathcal N}^3}
\frac{1}{\delta \Omega \, \delta \Omega'} \delta_{\theta,0} 
\delta_{\theta',0} 
\end{array}
\end{equation}

This can be rewritten in terms of the number of pairs,
$
n_p = \frac{1}{2}{\mathcal N}^2 \delta \Omega \Omega
$,
\eq
\label{nongaussw}
\begin{array}{ll}
\langle \hat{w}_{\Delta \theta}(\theta)\hat{w}_{\Delta \theta}(\theta') 
\rangle \;  -&\langle \hat{w}_{\Delta \theta}(\theta) \rangle \langle 
\hat{w}_{\Delta \theta}(\theta') \rangle
\\
&
=\frac{2}{\Omega}\int \frac{K d K}{2 \pi} J_{0,\Delta \theta}(K \theta) 
J_{0,\Delta \theta}(K \theta') [P_2^2(K) +\frac{2}{\mathcal N} P_2(K)] 
+\frac{1}{n_p}\delta_{\theta,\theta'} \\
&
 \; +\frac{1}{n_p}
[ w_{\Delta \theta} (\theta)
\delta_{\theta,\theta'} +
w_{\Delta \theta} (\theta) \delta_{\theta',0}
\frac{\delta \Omega} {\delta \Omega'} 
+w_{\Delta \theta} (\theta')\delta_{\theta,0}] \\
& \; +
\frac{1}{2 n_p}
\frac{\delta_{\theta,\theta'}}{{\mathcal N}\delta \Omega} \delta_{\theta,0} 
\end{array}
\end{equation}
and $\Omega = 4 \pi f_{sky}$.  This shows the often used $1/n_p^{1/2}$ shot 
noise error estimate
in the context of the full correlation function error matrix.

The Poisson rather than Gaussian contributions to the error matrix
are the terms on the last two lines above.  As these terms
are proportional to $1/n_p$ they will decrease more quickly
with increased binning than the terms involving Bessel functions,
i.e. the fractional contribution of the Poisson terms will
decrease as the binning increases (assuming the binned $w$ doesn't
increase).
The main new contribution is the term proportional to 
$\delta_{\theta,\theta'} w_\Delta(\theta)/n_p$ which
is significant only when $w(\theta)$ is "large" (relative to 
one and for large shot noise), i.e. it is most important
for small $\theta$.  The other
terms seem less significant as they only arise when either
$\theta$ or $\theta'$ are zero.

The terms analogous to the $(1+w)/n_p$ were found for the
angular correlation function calculated via the Landy-Szalay 
estimator (Landy \& Szalay \cite{LanSza93}) and extended by 
Bernstein \cite{Ber94} to the case where the correlation function
is large enough that the terms depending upon the power are also
important.  

For the power spectrum, the angle average gives
\eq
\label{poiimp}
\begin{array}{ll}
\langle \hat{P}_2(K) \hat{P}_2(K') \rangle -
\langle \hat{P}_2(K) \rangle \langle \hat{P}_2(K') \rangle 
&=
\int \frac{d \phi_k}{2 \pi}\frac{d \phi_k}{2 \pi}
\{(P_2(\bbK) + \frac{1}{\mathcal N})^2\frac{(2 \pi)^2}{\Omega} (
\delta_D({\bbK- \bbK'}) + \delta_D({\bbK+ \bbK'}) )
\\ & \; +
 \frac{1}{{\mathcal N}^2 \Omega}[2P_2(\bbK)+2P_2(\bbK')+P_2(|\bbK + \bbK'|) +
P_2(|\bbK - \bbK'|)]\\
& \;  + \frac{1}{{\mathcal N}^3 \Omega} \}
 \\
\\
&=\frac{2 \pi}{\Omega}\frac{2}{|K|}\delta_D(K-K')
(P_2(K) + \frac{1}{\mathcal N})^2 \\
& \;  +\frac{1}{{\mathcal N}^2 \Omega}[2P_2(K)+2P_2(K')
+4 \pi 
\int \theta d \theta J_0(K \theta) J_0(K' \theta) w(\theta)] 
\\ 
& \; +\frac{1}{{\mathcal N}^3 \Omega}\\
\end{array}
\end{equation}
 
The top line, diagonal in $K,K'$, is the usual contribution when
the shot noise is Gaussian, rather than Poisson.  
Here to connect with other formulae in the literature one can
make $K$ discrete\footnote{The discretization can be
done by binning, i.e. integrating 
from $k-1/2$ to $k+1/2$ for each value of $k$.
Assuming $k \gg 1$, below are shown only the first terms in 
$1/k$.}, in particular
$\delta_D(K-K')\rightarrow \delta_{K,K}/\Delta K = \delta_{K,K'}$.  Then,
replacing $K \rightarrow \ell, \delta_{K,K'}= \delta_{\ell,\ell'},
 P_2(K) \rightarrow C_\ell$ and $\Omega = 4 \pi f_{sky}$,
we can recognize the standard expression for large $\ell$
\eq
\langle \delta C_\ell \delta C_{\ell'} \rangle = 
\delta_{\ell, \ell'} \frac{2}{(2 \ell +1) f_{sky}}
(C_\ell + \frac{1}{\mathcal N})^2  \sim
\delta_{\ell, \ell'} \frac{1}{ f_{sky}} \frac{1}{\ell }
(C_\ell + \frac{1}{\mathcal N})^2
\end{equation}
The completely binned expression is found by replacing
$P_2(K)$ by 
\eq
P_2(K) \rightarrow \frac{1}{K \Delta K} \int_{-\Delta K/2}^{\Delta K/2}
d (\Delta \tilde{K}) (K+\Delta \tilde{K}) P_2(K + \Delta \tilde{K})
\end{equation} 
and likewise for $P_2(K')$
on the left hand side of eqn. \ref{poiimp} and averaging the
functions of $K,K'$ on the right hand side analogously.

The Poisson terms are non-diagonal and introduce
cross correlations in the power spectrum.
The cross correlation grows with
the separation $K-K_{ref}$ up to some limiting value, ignoring binning
one has
\eq
\frac{\langle P_2(K_{ref}) P_2(K) \rangle}{\sqrt{
\langle P_2(K_{ref})P_2(K_{ref}) \rangle \langle P_2(K) P_2(K) \rangle}}
\rightarrow_{K {\rm large}}
{\mathcal N}^{3/2}\frac{2 P_2(K_{ref})/{\mathcal N}^2 + 1/{\mathcal N}^3}
{\sqrt{\langle P_2(K_{ref})P_2(K_{ref}) \rangle}} \; ;
\end{equation}
This additional covariance makes the power spectrum errors correlated
even if the density distribution was originally Gaussian.  The size
of this cross correlation is shown in examples in section four,
it can be large.

The errors have both Gaussian and non-Gaussian contributions,
the non-Gaussian terms become comparable to the Gaussian terms when
$2 \pi {\mathcal N}^2 P_2(K)/K
\leq 1$ and $P_2(K) K \geq 2 \pi $, when there is no binning.
In contrast to the correlation function case, increasing
the binning 
increases the importance of the additional non-Gaussian terms
as binning roughly divides the first (i.e. Gaussian) term by
the bin size relative to the others.  In section four it
is shown that
for ${\mathcal N}$, $P_2(K)$ and $K$ of interest for SZ 
selected galaxy cluster
surveys the Poisson contribution to the errors can be relatively large.

\section{Three dimensions}
In three dimensions the correlation function $\xi(r)$ and power
spectrum $P(k)$ are related via
\eq
\begin{array}{l}
P(\bbk) =\int d^3 r e^{-i \bbk \cdot \bbr} \xi(\bbr) \\
\xi(\bbr) = \frac{1}{(2\pi)^3}\int d^3 k
e^{i \bbk \cdot \bbr} P(\bbk) \\
\end{array}
\end{equation}
To angle average these we again need to assume isotropy and 
that boundary effects aren't important.
In three dimensions, ${\mathcal N}$ denotes the number density per
$(h^{-1} {\rm Mpc})^3$.
The three dimensional continuum limit of the
Meiksin \& White 
\cite{MeiWhi99} error matrix calculation is
\eq
\label{3dperror}
\begin{array}{ll}
\langle \hat{P}(\bbK) \hat{P}(\bbK') \rangle -
\langle \hat{P}(\bbK) \rangle \langle \hat{P}(\bbK') \rangle 
&= \langle\hat{P}(\bbK)\hat{P}(\bbK')\rangle-P(\bbK)P(\bbK') \\
 &=\frac{(2 \pi)^3}{V}
(P(\bbK) + \frac{1}{\mathcal N})^2 (\delta_D(\bbK -\bbK') + 
\delta_D(\bbK +\bbK')) \\
&+ \frac{1}{{\mathcal N}^2 V}[ P(|\bbK + \bbK'|)
+ P(|\bbK-\bbK'|)+2 P(\bbK) +2 P(\bbK')] \\ 
& + \frac{1}{{\mathcal N}^3 V} \\
\end{array}
\end{equation}

Fourier transforming to get the correlation function, 
angle averaging via
\eq
\frac{1}{4 \pi} \int d \Omega \frac{1}{4 \pi} \int d \Omega'
\end{equation}
and using
\eq
\delta_D(\bbr-\bbr') = \frac{\delta_D(r-r')}{r^2} 
\delta_D(\Omega-\Omega')  
\end{equation}
gives
\eq
\begin{array}{ll}
\langle \hat{\xi}(r) \hat{\xi}(r') \rangle -
\langle \hat{\xi}(r) \rangle \langle \hat{\xi}(r') \rangle &=
\frac{1}{V \pi^2} \int k^2 dk (\frac{\sin(kr)}{kr}) (\frac{\sin(kr')}{kr'}) 
P(k)^2 \\
& \; +\frac{2}{V{\mathcal N} \pi^2 }\int k^2 dk (\frac{\sin(kr)}{kr}) 
(\frac{\sin(kr')}{kr'}) 
P(k)
+\frac{1}{V{\mathcal N}^2}\frac{2}{4 \pi r^2 } \delta_D(r-r') \\
& \; +\frac{1}{V{\mathcal N}^2}[\frac{2}{4 \pi r^2} \delta_D(r-r')
\xi(r)+ 2 \xi(r) \frac{1}{4 \pi r^{'2}}\delta_D(r') 
+ 2 \xi(r') \frac{1}{4 \pi r^{'2}}\delta_D(r)]
\\ 
& \; + \frac{1}{V{\mathcal N}^3}
\frac{1}{4 \pi r^2} \delta_D(r) 
\frac{1}{4 \pi r^{'2}}\delta_D(r') \} \\
\end{array}
\end{equation}
The additional non-Gaussian terms are the same as in the two
dimensional case, i.e. diagonal and proportional to $\xi(r)/n_p$ when binning
is included plus contributions if either $r$ or $r'$ are zero.
In this case $n_p = {\mathcal N}^2 V 4 \pi r^2 \Delta r /2$.
 
For the power spectrum, the angle average gives
\eq
\label{pow3a}
\begin{array}{ll}
\langle(\hat{P}(K) - P(K))(\hat{P}(K') - P(K')) \rangle
& =\frac{4 \pi^2}{|K|^2 V} 
(P(K) + \frac{1}{\mathcal N})^2 \delta_D(K-K') \\
&\; + \frac{2}{{\mathcal N}^2  V}(P(K) +P(K') + 4 \pi
\int r^2 dr \frac{\sin Kr}{Kr} \frac{\sin K'r}{K'r} \xi(r) ) \\
& \; + \frac{1}{{\mathcal N}^3  V} \\
\end{array}
\end{equation}

The error matrix for the power spectrum also has the same structure as in two
dimensions:  the Poisson terms introduce correlations between
different values of $K$, and
analogously, the non-Gaussian nature of the shot noise will
start to become important for the errors if 
$ 4 \pi^2 \frac{{\mathcal N}^2 P(k)}{|K|^2} \leq 1$ and
$|K|^2 P(K) > 4 \pi^2$.  With binning the first (Gaussian) term gets a factor
of one over the size of the bin, in three dimensions this coefficient
is likely to be less than one.

\section{Errors without Bessel functions}
The error matrices for the power spectrum and the correlation function
above each depend on both the correlation function and
the power spectrum.  This is less than ideal because
going between power spectra and correlation functions
involves integrals over Bessel functions $J_0$ 
in two dimensions and spherical Bessel functions $j_0(x) = \sin(x)/x$
in three dimensions.  Integrals against these Bessel functions do
not converge quickly; with real data and thus perhaps 
incomplete coverage in real or
momentum space the results may not be very accurate.\footnote{There are
fast ways of doing integrals against one Bessel function which
are described at http://casa.colorado.edu/$\sim$ajsh/FFTLog/ .}
If one wished to sidestep this problem by using the measured
correlation functions and power spectra together to calculate
the errors for either, one would be combining quantities which
have very different estimators and thus systematics.

However,
as the correlation function and power spectrum are 
Fourier transforms of each other, one can instead substitute the definitions
until one has errors in the correlation functions only in terms of
correlation functions or similarly for the power spectrum, integrated
against an integral of a combination of regular or spherical Bessel
functions.  This integral of (spherical) Bessel functions will
not depend on the power spectrum or correlation function and
can be calculated separately.
For three Bessel functions and three and four spherical
Bessel functions the exact value of this integral is
simple and can be used directly, for four Bessel functions the
result involves an elliptic integral.\footnote{The author will provide the
lengthy expression found, which is still partially in integral
form, upon request. The elliptic integral expression
is given in e.g. Watson \cite{Wat66}, page 414 or 
Van Deun \& Cools \cite{DeuCoo06}, page 594.}

This rewriting is only valid when the power 
spectrum or correlation function(s) is/are well behaved enough
that the order of integration can be changed.  In these cases it can
provide an immense simplification of the calculation.
The integral for three (spherical) Bessel functions in
two (three) dimensions is explicitly cut off, and the functions
involved are not as oscillatory as Bessel functions in all cases
where a simple rewriting was found,
providing more numerical stability for calculating the errors. 
Unless one uses the elliptic integral expression for the four Bessel
functions, this rewriting doesn't completely eliminate the Bessel 
functions needed to calculate the error matrix for (even 
a well behaved) $w(\theta)$.
The substitution and integrals are done below, first for the
two dimensional correlation function and power spectrum and then
their counterparts in three dimensions.  

For the two dimensional correlation function,
the error matrix for  $w$ is given by equation \ref{correr1}.
If the rise of $w(\theta)$ is shallow enough at the
origin, one can rewrite the term with just one $P_2(K)$ via
\eq
\begin{array}{l}
\int \frac{K d K}{2 \pi} J_0(K \theta) J_0(K \theta') P_2(K) \\
= \int \theta'' d \theta '' w(\theta'')
\int K d K  J_0(K \theta) J_0(K \theta') J_0(K \theta'') \\
= \int \theta'' d \theta '' w(\theta'') f(\theta,\theta',\theta'') \\
\end{array}
\end{equation}
where $f(\theta,\theta',\theta'')$
is known in closed form (Jackson \& Maximon \cite{JacMax72}),
i.e.
\eq
\label{fdef}
f(\theta,\theta',\theta'')=
\frac{\Delta(\theta,\theta',\theta'')}{2 \pi A(\theta,\theta',\theta'')} \; .
\end{equation}
Here 
\eq
\begin{array}{ll}
\Delta(\theta,\theta',\theta'') &= \left(
\begin{array}{ll}
1 \;& {\rm if } \; \theta+\theta'> \theta'', \theta+\theta''>\theta', 
\theta'+\theta'' > \theta \\
\frac{1}{2} \;& {\rm if } \; \theta+\theta'= \theta'' \; {\rm or} \;  
\theta+\theta''=\theta' \; {\rm or} \;
 \theta'+\theta'' = \theta \\
0 \;& {\rm otherwise} \\
\end{array} \right. \\
A(\theta,\theta',\theta'') &= \frac{1}{2}\sqrt{(\theta+\theta'+\theta'')
(-\theta+\theta'+\theta'')(\theta-\theta'+\theta'')(\theta+\theta'-\theta'')}
\end{array}
\end{equation}
Geometrically, $A$ is the area of the triangle formed with 3 lines of length
$\theta,\theta',\theta''$; $\Delta$ is 1 if the triangle is
non-degenerate, 1/2 if it is degenerate and zero if no triangle is
formed.  In particular, $\Delta$ will cut off the integral over $\theta''$
for $\theta''>\theta + \theta'$.  For any two sides equalling the
third $f$ diverges, but in the integral it is an integrable divergence;
integrating by parts gives a finite result.

Unfortunately a simple expression (without elliptic integrals) for
\eq
\int dk \,  k J_0(k \theta) J_0(k \theta') J_0(k \theta'') J_0 (k \theta''')
\end{equation}
was not found,
so there is still some explicit $P_2(K)$
dependence in $\langle w w \rangle$.  

In terms of $f(\theta,\theta',\theta'')$:
\eq
\label{correr2}
\begin{array}{ll}
\langle \hat{w}(\theta)\hat{w}(\theta') \rangle
-\langle \hat{w}(\theta) \rangle \langle \hat{w}(\theta')) \rangle
&=\frac{2}{\Omega}\int \frac{K d K}{2 \pi} J_0(K \theta) J_0(K \theta') P_2^2(K)
+\frac{4}{\Omega {\mathcal N}}
 \int d \theta '' \theta '' w(\theta'') f(\theta,\theta',\theta'') \\
&+\frac{1}{\Omega {\mathcal N}^2}
[2 (1+w(\theta))
\frac{\delta_D(\theta-\theta')}{2 \pi \theta} +
 2 w(\theta) \frac{\delta_D(\theta')}{2 \pi \theta'} 
+2 w(\theta') \frac{\delta_D(\theta)}{2 \pi \theta}] \\
& +
\frac{1}{\Omega {\mathcal N}^3}
\frac{1}{(2 \pi \theta)(2 \pi \theta') } \delta_D(\theta)
\delta_D(\theta')
\end{array}
\end{equation}

For the angular power spectrum, there is no quadratic term in $w$ so
if $P_2(K)$ is not too singular at the origin, its error matrix can be
rewritten completely in terms of $P_2(K)$ itself:
\eq
\begin{array}{ll}
\langle \hat{P}_2(K) \hat{P}_2(K') \rangle -
\langle \hat{P}_2(K) \rangle \langle \hat{P}_2(K') \rangle 
&=\frac{2 \pi}{\Omega}\frac{2}{|K|}\delta_D(K-K')
(P_2(K) + \frac{1}{\mathcal N})^2  \\
&+ \frac{1}{{\mathcal N}^2 \Omega}(2P_2(K)+2P_2(K')) 
+
\frac{2}{{\mathcal N}^2 \Omega}
\int K'' \, d K'' f(K,K',K'')P_2(K'') 
\\
&+\frac{1}{{\mathcal N}^3 \Omega}\\
\end{array}
\end{equation}
where $f(K,K',K'')$ is the same function as in equation
\ref{fdef}, with the replacement $\theta \to K$ etc.
Smoothing is done straightforwardly.

For three dimensions all the integrals over spherical Bessel functions 
can be done exactly, and the applicability of this substitution is limited
only by the requirement of sufficient falloff of the power spectrum and 
the correlation function at the origin.
In terms of these integrals the error matrices from section three become
\eq
\begin{array}{ll}
\langle \hat{\xi}(r) \hat{\xi}(r') \rangle -
\langle \hat{\xi}(r) \rangle \langle \hat{\xi}(r') \rangle &=
\frac{16}{V}
\int r^{''2} dr'' r^{'''2} dr '''
\xi(r'') \xi(r''')  g_3(r,r',r'',r''')\\
&+\frac{8}{V{\mathcal N} \pi}
\int r^{''2} dr'' \xi(r'') f_3(r,r',r'') 
 \\
&+\frac{1}{V{\mathcal N}^2}[\frac{2}{4 \pi r^2} \delta_D(r-r')
(1 + \xi(r))+ 2 \xi(r) \frac{1}{4 \pi r^{'2}}\delta_D(r') 
+ 2 \xi(r') \frac{1}{4 \pi r^{'2}}\delta_D(r)]\\ &+ \frac{1}{V{\mathcal N}^3}
\frac{1}{4 \pi r^2} \delta_D(r) 
\frac{1}{4 \pi r^{'2}}\delta_D(r') \} \\
\end{array}
\end{equation}
for the correlation function
and
\eq
\label{powf3}
\begin{array}{ll}
\langle(\hat{P}(K) - P(K))(\hat{P}(K') - P(K')) \rangle
& =\frac{4 \pi^2}{|K|^2 V} 
(P(K) + \frac{1}{\mathcal N})^2 \delta_D(K-K') \\
&\; + \frac{2}{{\mathcal N}^2  V}(P(K) +P(K') +
\frac{2}{\pi} \int_{0}^{\infty} dK'' K^{''2} P(K'')f_3(K,K',K'')) \\
& \; + \frac{1}{{\mathcal N}^3  V} \;  \\
\end{array}
\end{equation}
for the power spectrum.

Here
\eq
\begin{array}{l}
f_3(r,r',r'') = 
\int_0^\infty k^2 dk \frac{\sin(k r)}{k r} \frac{\sin(k r')}{k r'} 
\frac{\sin(k r'')}{k r''} \\
 \; \; \; =
\frac{\pi}{8}
(-\sgn(r-r'-r'') + \sgn(r+r'-r'') +\sgn(r-r'+r'') -\sgn(r+r'+r'')) \\
 \; \; \; =
\frac{\pi}{8}
(-\sgn(r-r'-r'') + \sgn(r+r'-r'') +\sgn(r-r'+r'') -1)
\\
\\
g_3(r,r',r'',r''')=
\int_0^\infty k^2 dk \frac{\sin(k r)}{k r} \frac{\sin(k r')}{k r'} 
\frac{\sin(k r'')}{k r''} \frac{\sin(k r''')}{k r'''} 
\\ \; \; \; \; =
 \frac{\pi}{16} \{ |r -r'- r'' - r'''|-|r +r'- r'' - r'''|
 - |r -r'+ r'' - r'''| +|r +r'+ r'' - r'''|
\\ 
\; \; \; \; -  |r -r'- r'' + r'''|+ |r +r'- r'' + r'''|+            
     |r -r'+ r'' + r'''| -(r +r'+ r'' + r''') \}
\\
\end{array}
\end{equation}

As mentioned above,
these rewritings have many advantages besides getting rid of the
instabilities due to oscillations of the (spherical) Bessel functions.
The functions $f$ and $f_3$  cut off the integrals for large 
argument: i.e.
in two dimensions $f(\theta,\theta',\theta'')$ is zero when
$\theta'' > \theta+\theta'$  
and 
in three dimensions, e.g., for $K = K'$ in the power spectrum, the
term with $f_3$ becomes
\eq
\frac{2}{\pi}
 \int_{0}^{\infty}dK'' K^{''2} P(K'')
\frac{\pi}{8}(1 + \sgn(2K-K'')) = \frac{1}{2}
 \int_0^{2K} dK''K^{''2} P(K'')
\end{equation}
with an analogue that can be seen by inspection for $K \ne K'$.
These cutoffs reduce the dependence on the asymptotic values of the power
spectrum or the correlation function.
The function $g_3(r,r',r",r''')$
also goes to zero as $r''$ or $r'''$ goes to infinity, but as both
parameters are varying some care must be taken in the joint limits.

\section{Examples}
In this section the full Poisson error matrices given above 
are calculated for a power law power spectrum in two
and three dimensions and compared to the usual Gaussian error matrix.  
In addition, the effects on the angular power spectrum
errors for SZ selected galaxy clusters are shown.
For these examples the corrections are significant except for the two 
dimensional correlation function.
The signal is taken to be Gaussian; if
not, the three and four point functions mentioned earlier would need to
be included. 

Starting with two dimensions, consider a power law power spectrum
\eq
P_2(K) = A_k k^n
\end{equation}
with corresponding correlation function
$w(\theta) = A_\theta \theta^{-n-2} $.
For the error matrix to converge, $-1< n< -1/2$,
take $n = -0.9$ for illustration, $A_k = 5.0 \times 10^{-4}$
and thus $A_\theta = 7.0 \times 10^{-5}$.
This is a rough fit to the power spectrum for SZ selected
clusters for $Y_{min} = 1.7 \times 10^{-5}$ such as might be seen 
with APEX\footnote{http://bolo.berkeley.edu/apexsz/index.html}.
We vary the
number density of objects per steradian, ${\mathcal N}$ and the
binnings $\Delta \theta$ and $\Delta K$.
The objects are taken to have a  Gaussian distribution,
and $f_{sky} = 1/4 \pi$, i.e. one steradian.  The terms in
the error matrix all increase by a 
multiplicative
factor $1/f_{sky}$ for other values of $f_{sky}$.

The true cluster case differs from this in three ways: $P_2(K)$ and 
$w(\theta)$ are not true power laws, the number density is fixed 
(for
clusters with the above $Y_{min}$, ${\mathcal N} \sim 33000$ is expected) 
and the cluster distribution
is not expected to be Gaussian to arbitrarily small scales.
The cluster power spectrum
and number densities for an experiment such as APEX or 
Planck\footnote{http://astro.estec.esa.nl/SA-general/Projects/Planck} 
are shown later on, however the signal is still taken to be Gaussian and
$f_{sky}$ remains $1/4 \pi$.

For terminology, error means the square root of the diagonal (i.e. $K=K'$
for power) term in the error matrix.
The ``shot noise'' error means the term $n_p^{-1/2}$ in the
error of $w(\theta)$ and the term proportional to
$\frac{1}{|K|^{1/2}{\mathcal N}}$
in the error of $P_2(K)$.  The ``Gaussian error'' refers to the terms
which arise when the shot noise is considered to be Gaussian rather
than Poisson, i.e. the first line on the right 
in equations \ref{nongaussw} and
\ref{poiimp} and their three dimensional generalizations.
The ``non-Gaussian'' error or ``Poisson'' error
is everything else in the error.  Similarly, ``non-Gaussian''
and ``Poisson'' refer to the analogous terms in the rest of the error matrix.

Figure one shows the effects for the correlation function.  The delta
function errors which only appear at the origin have not been included.
\begin{figure*}[htb]
\label{fig1}
\begin{center}
\resizebox{7in}{!}{\includegraphics{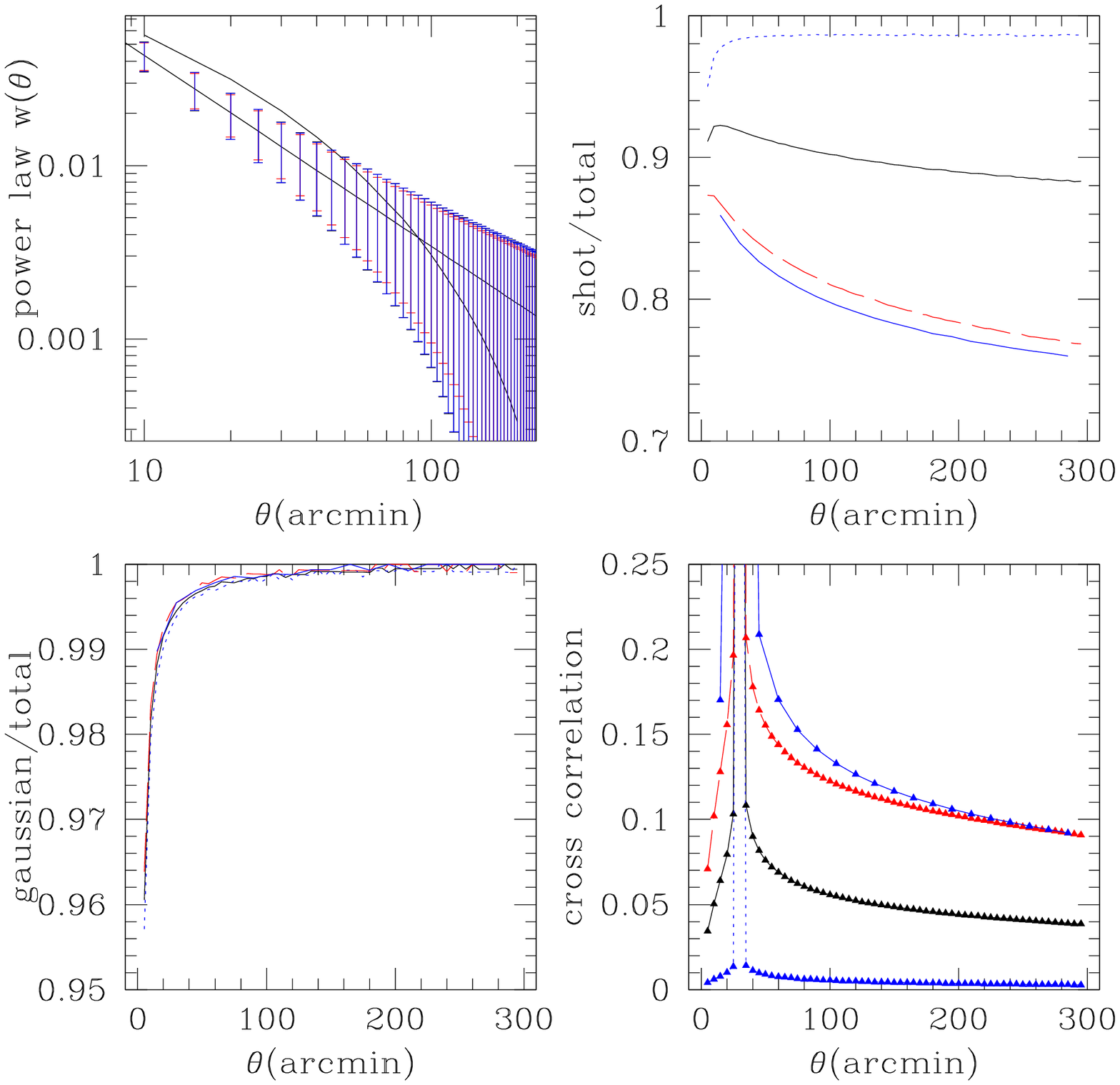}}
\end{center}
\caption{Upper left: Shot noise, Gaussian and full errors
for $w(\theta)$, with ${\mathcal N} = 35000$, 5 arcmin smoothing, 
and $f_{sky} = 1/4 \pi$.
The curved line is the angular power spectrum for SZ selected galaxy
clusters with $Y_{min} = 1.7 \times 10^{-5}$ as mentioned in the
text.
Upper right: shot noise as fraction of total correlation
function error for 5 arcmin smoothing and 
${\mathcal N}$ = 5000, 35000 and 65000 objects per steradian, and then
for 15 arcmin smoothing and ${\mathcal N = 35000}$, top
to bottom.  Lower left: the ratio of using only errors calculated with
the
Gaussian approximation for Poisson noise, 
divided by the full errors, for all four cases.
Lower right:  The cross correlation,
$\langle w(\theta) w(30') \rangle/\sqrt{
\langle w(30') w(30') \rangle/\langle w(\theta) w(\theta) \rangle}$
as a function of angle for the angular correlation function
for 5 arcmin smoothing and 
three different cases of object numbers (5000,35000,65000 from
bottom to top) and for 15 arcmin smoothing and ${\mathcal N} =
35000$ (very top).}
\end{figure*} 
At top left is the angular correlation function for ${\mathcal N} = 35000$
with shot, Gaussian and full error bars.
The smoothing is 5 arcmin, i.e. the spacing between the error bars.
The curved line is the correlation function for
the analogous SZ cluster sample with
$Y_{min} = 1.7 \times 10^{-5}$.  
Figure 1 upper right and lower left
show the fraction of shot noise alone to the total error
and Gaussian error to total error
as a function of changing ${\mathcal N}$ respectively.
As can be seen, although the shot noise is a significant source of
the error, using only the shot noise underestimates the
error by some noticeable fraction unless the shot noise is
extremely large (${\mathcal N} = 5000$ here).  The
non-Gaussian terms in the error are an equally small proportion
of the total error for
all three values of ${\mathcal N}$ considered here. 
The dominance of the shot noise in the error means that the cross
correlation decreases as the shot noise increases, 
as can be seen in the lower right hand panel for
the three different ${\mathcal N}$ values 
and also the case with binning of 15 arcmin rather than 5 arcmin.
As mentioned earlier, increased binning reduces the relative contribution
of the shot noise error to the total error and increases the cross correlation
for a fixed number density.

For the angular power spectrum, the corresponding quantities are
shown in figure two.  Here smoothing is in equal
intervals in log $K$.  The curved line is the
expected angular power spectrum for the SZ selected cluster
survey mentioned above.  The non-Gaussian contribution to 
the error matrix is very important!
\begin{figure*}[tbh]
\label{fig2}
\begin{center}
\resizebox{7in}{!}{\includegraphics{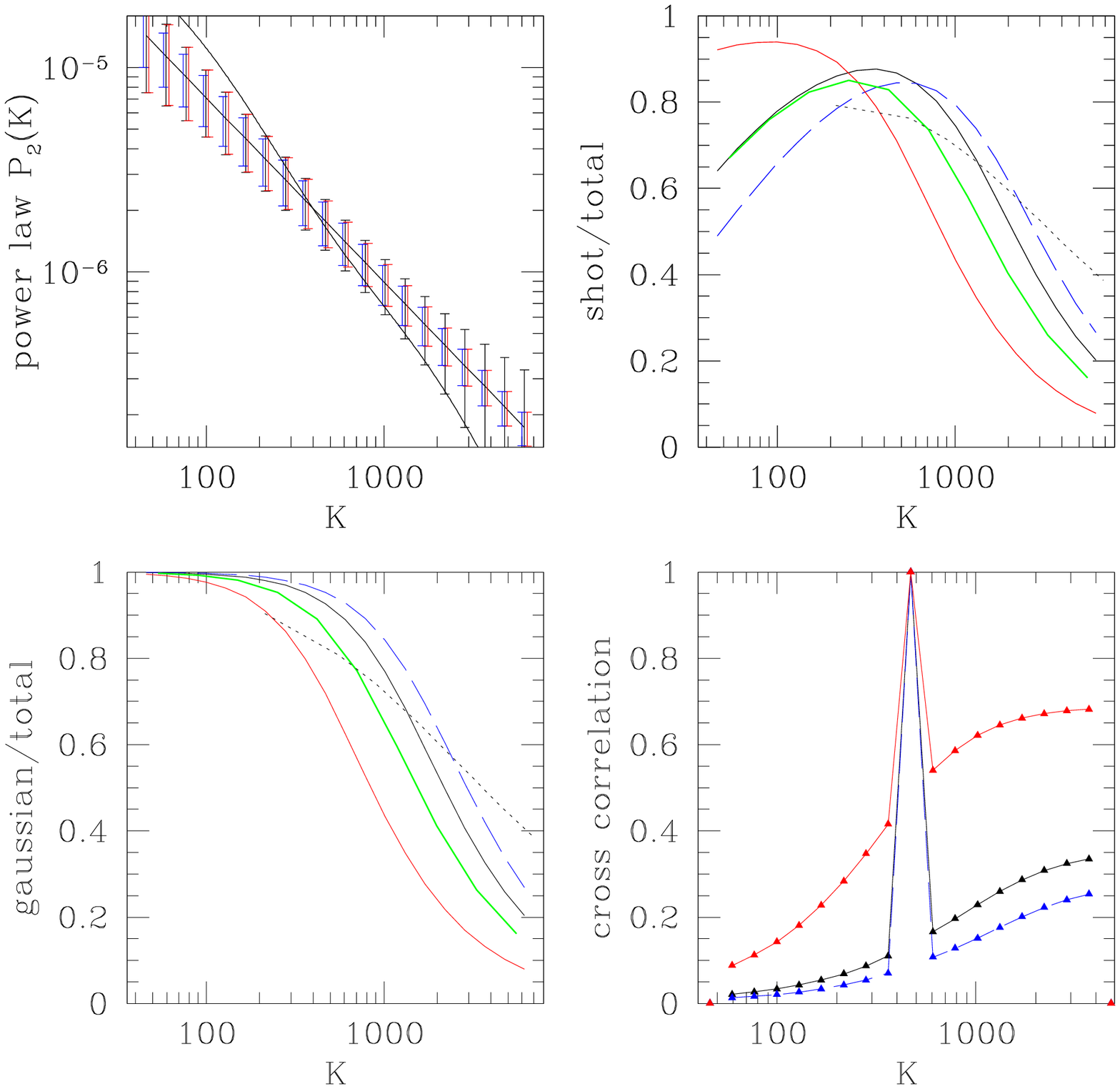}}
\end{center}
\caption{Upper left:  angular power spectrum with full errors, 
Gaussian errors, and only shot noise errors, 
largest to smallest error bars respectively. Error bars are slightly offset.
Upper right: shot noise only as fraction of total error for 20 K bins
smoothed in log K and 
${\mathcal N}$ = 5000, 35000 and 65000, and 
for 10 k bins and 20 equal $K$ rather than log $K$
bins with ${\mathcal N}$ = 35000.  
Lower left:  Gaussian errors divided by the full errors, for all five cases.  
The lowest line (at far right of each figure)
is for ${\mathcal N} = 5000$, the second lowest is that
for using 10 bins, and then the next two going up are
${\mathcal N} = 35000,65000$.  The highest (dotted) line 
is for 20 equal bins in $K$ rather than log $K$.
Lower right:  the cross correlation for the power spectrum,
$\langle P_2(469) P_2(K) \rangle/\sqrt{
\langle P_2(469)P_2(469) \rangle/\langle P_2(K) P_2(K) \rangle}$
as a function of $K$ for 20 log $K$ bins and three different
choices of object numbers (5000,35000,65000 from
top to bottom).}
\end{figure*} 
At upper left are the shot only, Gaussian and full errors
for ${\mathcal N} = 35000$, the curved line is
the reference SZ selected galaxy cluster
angular power spectrum mentioned earlier.
Upper right and lower left are the ratios of the shot noise only
and Gaussian errors only to the total error as a function of $K$,
and lower right is the cross correlation.  The cross
correlation is large and is entirely induced by the
Poisson nature of the shot noise given our assumptions that
$\eta$ and $\zeta$ are zero.
The effect of the binning, as $K>1$,
is to make the Gaussian contribution even a smaller fraction
of the total error matrix.  Binning in equal size bins in $K$
rather than log $K$ increases the Gaussian contribution to the
error at larger $K$ (as the bins are smaller
at large $K$ in this case), however the Poisson nature of the shot noise
still makes the ratio of the Gaussian to full error small. 
The case for
${\mathcal N} = 35000$ and 20 equal bins in $K$ rather than
log $K$ is shown to illustrate this. 
The change induced by using the Poisson rather than Gaussian
error matrix is large, for example, for $K=2000$ and ${\mathcal N} = 35000$
the error bar increases by close to a factor of two, and the
cross correlation with $K=469$ is  $\sim 0.5$ rather than zero.

The case of SZ selected galaxy clusters with specific modeling and
cosmological assumptions can be done analogously: i.e. with
the Evrard mass function (Evrard et al \cite{Evretal02}),
Sheth-Tormen bias (Sheth \& Tormen \cite{SheTor99}), 
Eisenstein-Hu (Eisenstein \& Hu \cite{EisHu97})
transfer function, $T_*^{SZ} = 1.2 keV$ and $\sigma_8 = 0.9$, 
$\Omega_m = 0.3$.  Details of the calculation to get this curve
and assumptions can be found in
Cohn \& Kadota \cite{CohKad04}.  The error matrix for
$Y_{min} = 1.7 \times 10^{-5}$ is 
extremely close to those for the ${\mathcal N} = 35000$ example in
figure one.
Figure three shows
the fraction of total errors given by the Gaussian approximation, and the 
cross correlation for $Y_{min} = 1.7 \times 10^{-5},
3.4 \times 10^{-5}, 1.0 \times 10^{-4}$ (with ${\mathcal N}$ determined
by $Y_{min}$ and the other parameters mentioned before). 
A $Y_{min} = 3.4 \times 10^{-5}$ might 
arise if the survey was wider and shallower, and $Y_{min} = 10^{-4}$ is
a rough estimate of the errors for Planck (even smaller effective
$Y_{min}$ might arise once cluster finding is applied (Geisbusch et
al \cite{GeiKneHob04}, for example).
\begin{figure}[tbh]
\label{fig3}
\begin{center}
\resizebox{3.5in}{!}{\includegraphics{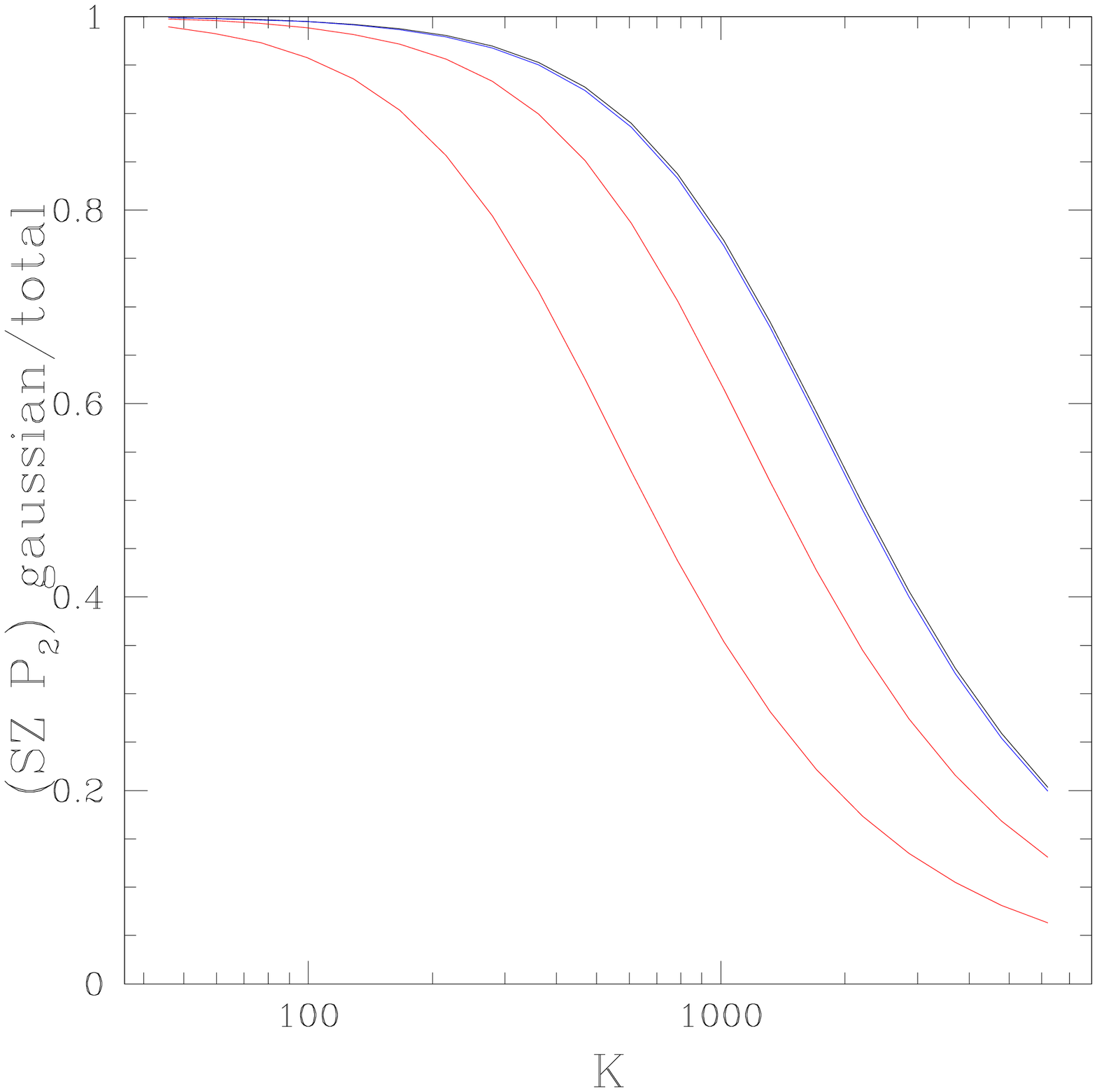}}
\resizebox{3.5in}{!}{\includegraphics{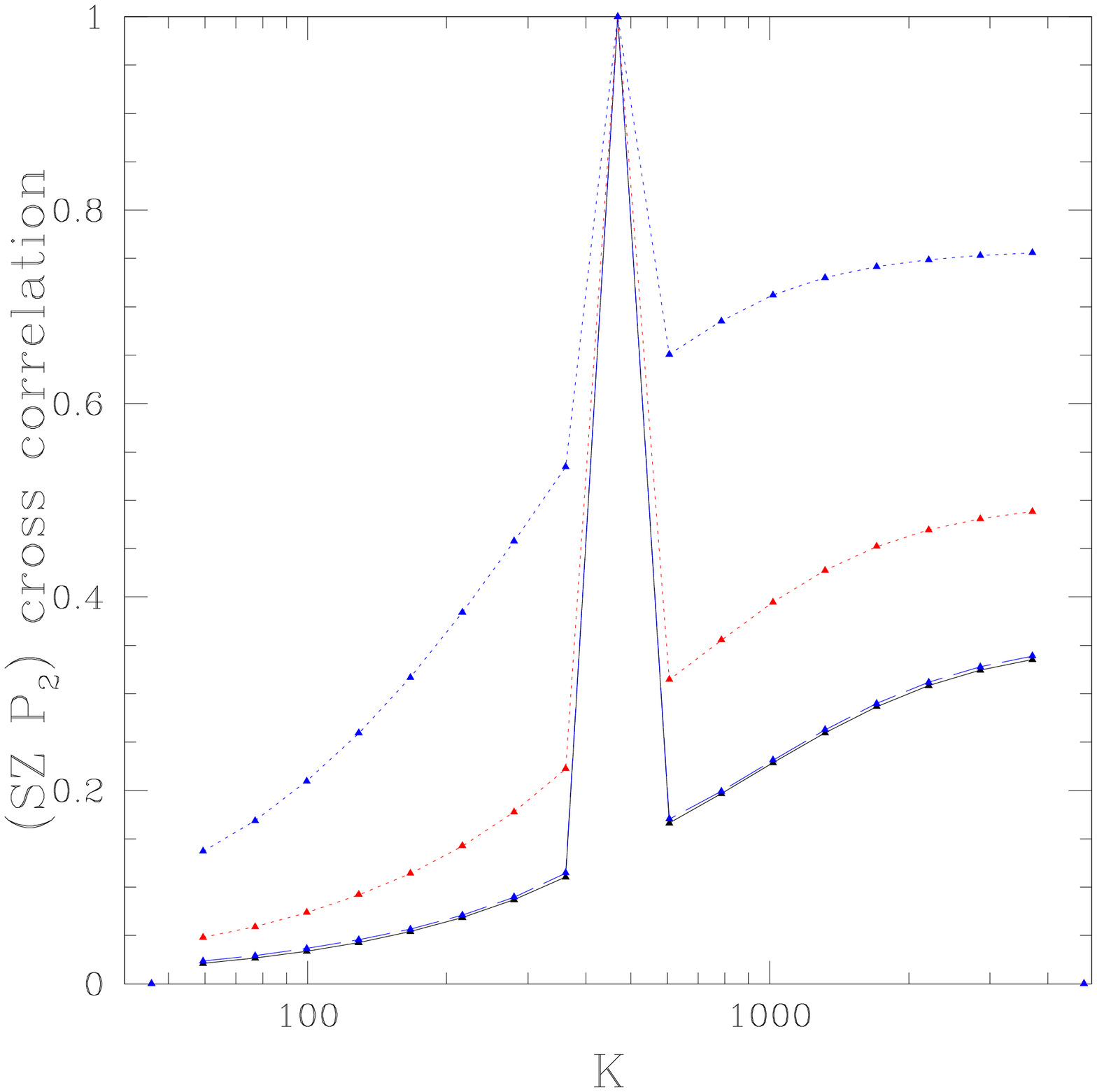}}
\end{center}
\caption{Top: the ratio between the Gaussian and full error for
three different choices of $Y_{min}$,
$1.7 \times 10^{-5}, 3.4 \times 10^{-5}, 1.0 \times 10^{-4}$;
top to bottom.  
Bottom:  the cross correlation for the power spectrum,
$\langle P_2(469) P_2(K) \rangle/\sqrt{
\langle P_2(469)P_2(469) \rangle/\langle P_2(K) P_2(K) \rangle}$ 
for
three different choices of $Y_{min}$,
$1.7 \times 10^{-5}, 3.4 \times 10^{-5}, 1.0 \times 10^{-4}$, bottom to top.
In both figures there is also a line for for the power law example 
with ${\mathcal N} = 35000$; it is almost completely degenerate with that
for $Y_{min} = 1.7 \times 10^{-5}$.}
\end{figure}
The above is for the power spectrum.  For the correlation function, the
Poisson corrections are small, similar to the power law example.
For instance, in estimates pertaining to Planck (e.g. Mei \& Bartlett 
\cite{MeiBar03}) with $Y_{min} = 1.0 \times 10^{-4}$, the shot noise only
approximation to the error is quite good, within the 5\% of 
the full Poisson error for $\theta > 30'$. 

In three dimensions the value of $K$  for clusters is smaller and
$\xi(r)$ is larger, so the Gaussian terms become more dominant for the 
power spectrum and less so for the correlation function.   
An unbinned power law case is shown here just to illustrate the 
(large) effects of the full Poisson treatment.
Consider a rough power law approximation to
the correlation function for galaxy clusters (e.g. Bahcall et 
al \cite{Bahetal}, for $M > 1 \times 10^{14} h^{-1} M_\odot$), i.e.
take $\xi(r) = 100 r^{-1.9}$ so that $P(k) = 1.9 \times 10^{3} k^{-1.1} 
(h^{-1} Mpc)^3$. 
In figure four, the ratio of the Gaussian to total errors and the 
cross correlation are shown for
the correlation function and the power spectrum,
with ${\mathcal N}$ = $10^{-3}, 10^{-4},10^{-5}$.
For the correlation function the Gaussian errors
differ from the Poisson errors by a factor of $\xi(r)/n_p$ and
are a less accurate approximation for the correlation function error
than in two dimensions.
\begin{figure*}[tbh]
\label{fig4}
\begin{center}
\resizebox{7in}{!}{\includegraphics{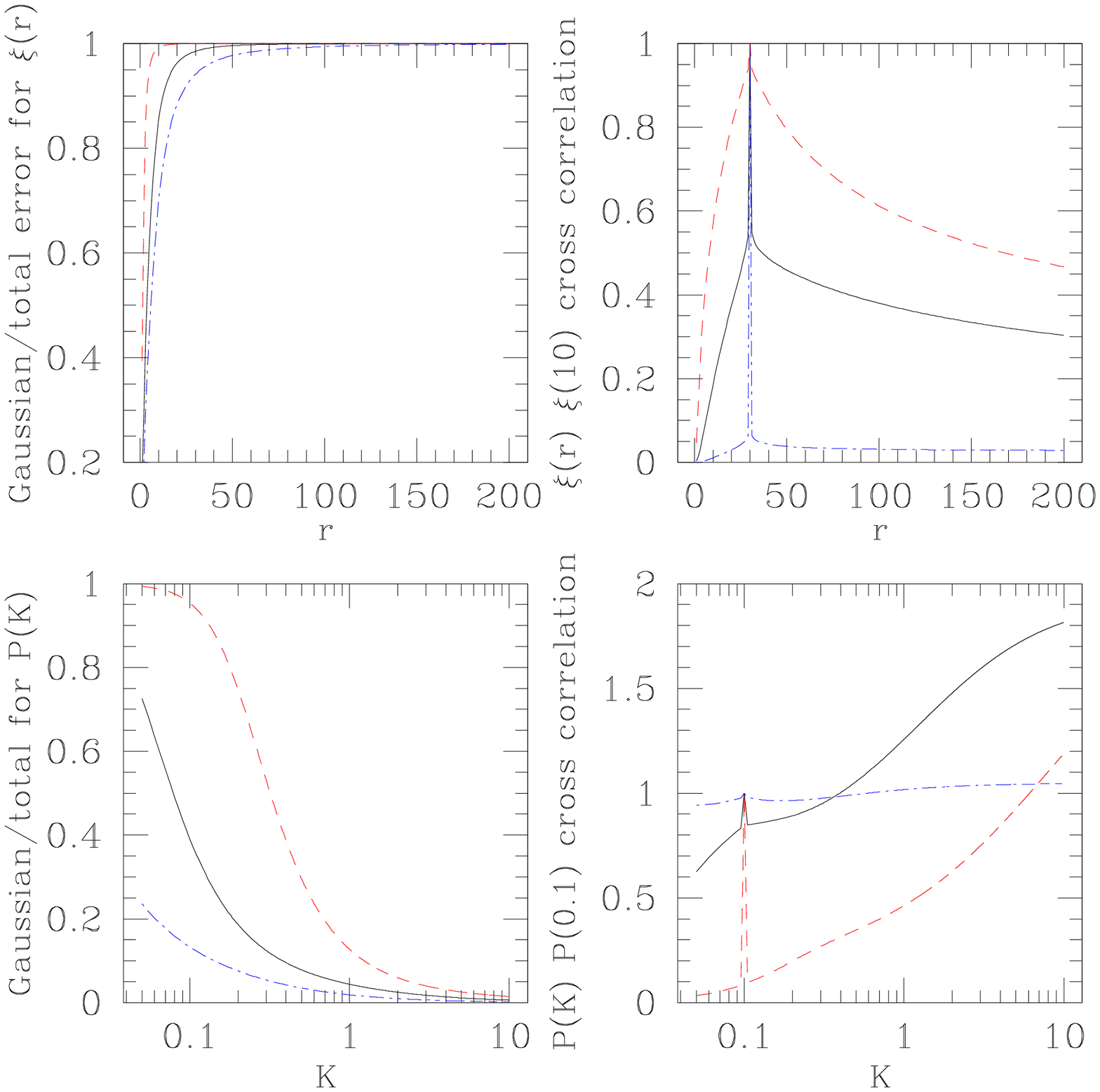}}
\end{center}
\caption{Upper left:
the ratio of the Gaussian error to the
total error for the three dimensional correlation function.
Upper right:
the cross correlation  $\langle \xi(r) \xi(30) \rangle
/\sqrt{\langle \xi(r) \xi(r) \rangle \langle \xi(30) \xi(30) \rangle}$
for the three dimensional power law example.  For both,
${\mathcal N} = 10^{-3},10^{-4},10^{-5}$ top to bottom.
Lower left: the ratio of the Gaussian error to the
total error and right, the cross correlation  $\langle P(k)P(0.1) \rangle
/\sqrt{\langle P(k)P(k) \rangle \langle P(0.1) P(0.1) \rangle}$
for the three dimensional power law example.
${\mathcal N} = 10^{-3},10^{-4},10^{-5}$ top to bottom left, and bottom to top
right. No binning is assumed, so the peak in the cross
correlation is very sharp.}
\end{figure*}
One expects of order $10^{-4}-10^{-5}$ clusters,
with mass above $10^{14} h^{-1} M_\odot$ per cubic $h^{-1} Mpc$; if the 
mass cut is higher the density is lower and the shot noise effects
and the correlation function/power spectrum are all larger.
The Poisson effects on the error matrix
appear to be large unless the shot noise is very small.

For both two and three dimensions, an estimate of when non-Gaussian
effects in the (angular) power spectrum itself become important is often
given by considering $\Delta_2^2(k) = \frac{K^2 P_2(K)}{2 \pi}$ and
$\Delta^2(K) = \frac{K^3 P(K)}{2 \pi^2}$ in the range $\sim 0.1-1$.
For the power law examples considered this is a large range,
$650 \leq K \leq 5300$ for two dimensions and $0.03 \leq K \leq 0.09$
for three dimensions, i.e. there may be other modifications
at the scales where the shot noise non-Gaussianity becomes relevant.

\section{Conclusions}
Poisson shot noise can produce an error matrix for the correlation
function and power spectrum significantly different from that calculated in
the Gaussian shot noise approximation.  The primary
contribution to the correlation function error matrix is diagonal and
goes as $w(\theta)/n_p$ in two dimensions and as
$\xi(r)/n_p$ in three dimensions.  For the power spectrum, there are
additions to both diagonal and off-diagonal terms in the error matrix, and
thus correlations are introduced in the power spectrum, even
for a Gaussian density distribution.  For SZ selected galaxy clusters, 
the errors
in the two dimensional correlation function are well approximated by
the Gaussian error except for $\theta$ extremely small.
For the power spectrum, the cross correlations
and additional errors are significant.  Binning increases this effect
as it reduces the Gaussian contribution relative to the 
Poisson contributions for the power spectrum.
This increase in the error matrix due to the Poisson nature of shot noise 
should be taken into consideration both for parameter estimation
and in survey design (e.g. a shallow survey increases the shot noise
and thus these additions to the error and correlations for 
the power spectrum).

There are two important caveats:
The cluster density itself is non-Gaussian at
small enough scales and thus at small enough distances
these two contributions to the error matrix should be combined.
It should be noted that due to non-Gaussianity, the error matrix alone
will also be insufficient to calculate the full likelihood.\footnote{It
is an interesting question when the error matrix is sufficient for
calculating the full likelihood and how that compares to when the
non-Gaussianity of the shot noise becomes important.  I thank the
referee for raising this issue.} 
The Poisson shot noise description may also break
down in high density regimes (Casas-Miranda et al \cite{Casetal02}).
It would be interesting to see the effects of the other non-Gaussian
shot noise distributions which appeared in their analysis (sub- and
super-Poisson).  

The analytic calculations here, just as in the Gaussian shot noise case, 
are of the most utility in estimating the power of future observations and
guiding the corresponding observational strategies.
When the eagerly awaited data is in hand, mock catalogues and various
statistical strategies such as bootstrap, jack-knife and Monte Carlo
(whichever is most appropriate for the question of interest, see e.g.
Lupton \cite{Lup93} for an introduction) will
likely be needed to include the effects of both the complications 
discussed in this paper and additional observational aspects such
as window functions.  

An additional point in this paper is that there is a rewriting of integrals
used in the error matrix which 
eliminates most of the integrals over Bessel and spherical Bessel functions. 
This reduces much of the dependence of the error matrix on the tails of the 
power spectrum and correlation functions and also makes the error matrix
integrals more tractable.

Acknowledgements:  This work was supported in part by NSF-AST-0205935.
I thank W. Hu for suggesting the inclusion of Poisson 
shot noise for galaxy cluster errors.
I also thank T.-C. Chang, G. Holder, 
Y. Lithwick, S. Mei, A. Pope, R. Scranton, R. Sheth, I. Szapudi, R. 
Wechsler, the anonymous referees, and especially
M. White for helpful discussions.  I thank R. Sheth for correcting errors
in some of the integration measures.

\end{document}